\documentclass{aa}  

\usepackage{color}
\usepackage{float}
\usepackage{soul}
\usepackage{dblfloatfix}
\usepackage{hyperref}
\hypersetup{
    colorlinks=true,
    citecolor = blue,
    linkcolor=blue
            }
\usepackage{graphicx}
\usepackage{longtable}
\usepackage{multicol}
\usepackage{supertabular}

\usepackage{txfonts}
\usepackage{natbib}
\newcommand{\cahk}{Ca\,{\sc ii}\,H\&K}
\newcommand{\kms}{\,kms$^{-1}$}
\usepackage[dvipsnames]{xcolor}%
\begin{document}

   \title{Four decades of photometry of XX\,Trianguli, `the most spotted star' in the sky}

\author{Zs.~K\H{o}v\'ari 
          \inst{1,2}
        \and
          K.~G.~Strassmeier
          \inst{3,4}
        \and
          K.~Ol\'ah 
          \inst{1,2}
        \and
          B. Seli
          \inst{1,2}
        \and
          G.W. Henry
          \inst{5}
        \and 
          K. Vida  
          \inst{1,2}
  }

   \institute{Konkoly Observatory, HUN-REN Research Centre for Astronomy and Earth Sciences, Konkoly Thege \'ut 15-17., H-1121 Budapest, Hungary\\
              \email{kovari@konkoly.hu}
        \and
        HUN-REN RCAES, MTA Centre of Excellence, Budapest, Konkoly Thege út 15-17., H-1121 Budapest, Hungary
        \and
        Leibniz-Institute for Astrophysics Potsdam (AIP), An der Sternwarte 16, D-14482 Potsdam, Germany
        \and
        Institute for Physics and Astronomy, University of Potsdam, Karl-Liebknecht-Strasse 24/25, D-14476 Potsdam, Germany
        \and
        Tennessee State University, Nashville, TN 37209, USA (retired)
            }

   \date{Received ...; accepted ...}

 
  \abstract
   {Over the past 40 years the brightness variations of XX\,Tri, a single line RS\,CVn type binary system with a synchronized K-giant primary, has exceeded one magnitude in the $V$ band. Although these changes are primarily caused by starspots, an additional activity-related mechanism may also be behind the long-term trend of overall brightness increase.}
   {By compiling the most complete photometric data set so far, we attempt to examine how the nature of seasonal-to-decadal changes can be linked to global magnetism.}
   {To find the long-term activity cycles and their properties a time-frequency analyser code is used. We divided the entire data set into four consecutive intervals, for which we separately determined the Fourier spectra around the orbital period. This is performed with a Fourier-transformation based frequency analyser tool.}
   {The long-term brightening of XX\,Tri was accompanied by a gradual increase in the effective temperature, which resulted in a blueing shift in the Herzsprung-Russell diagram. In the long term, a constant cycle of about 4 years is most strongly present in the entire data. Besides, we also found a modulation of about 11 years, and a slowly decreasing cycle of about 5.7--5.2 years. 
   From the seasonal datasets we found that the most dominant rotation-related periods are scattered around the orbital period. From this we infer a solar-type surface differential rotation, although the surface shear is significantly smaller than that of the Sun.}
   {The 4-year cycle indicates flip-flop-like behavior: during this time, the 2--3 active longitudes usually present on the stellar surface are rearranged. The magnitude-range changes in the long term cannot be interpreted solely as changes in the number and size of spots; the unspotted brightness of XX\,Tri has also increased over the decades. This should alert users of photometric spot models to reconsider the basic concept of constant unspotted brightness in similar cases.}


   \maketitle
%

\section{Introduction}

Dark starspots, first mentioned by \citet{1947PASP...59..261K}, are a long-known feature of cool stars with convective envelopes. A characteristic subgroup of these stars is RS\,Canum Venaticorum (RS\,CVn) variables, which are close binaries with a cool component showing chromospheric activity (e.g., starspots). Behind the magnetic activity is the dynamo mechanism, which is essentially activated by the convection of conductive plasma and the rotation of the star. Fast rotation is one of the important conditions for amplifying magnetic activity. In the case of RS\,CVn type stars with an active cool giant component, the binarity plays a prominent role, namely that in such close binary systems, due to the tidal synchronization, the rotation of the star remains fast, even though blowing up to a giant star would just slow down the rotation. However, tidal forces in such systems not only help maintain fast rotation, but tidal interactions are supposed to modify the operation of magnetic dynamo itself \citep{2003A&A...405..291H,2003A&A...405..303H}. 
Such a manifestation is, for example, when active longitudes form at certain phases fixed to the orbit \citep[e.g.,][]{2006Ap&SS.304..145O}. On the other hand, in RS\,CVn binaries, the magnetic activity can feed back on the orbital dynamics, causing orbital period changes on the timescale of the activity cycle, as described by the Applegate mechanism \citep{1987ApJ...322L..99A,1992ApJ...385..621A}.
Therefore, it is no exaggeration to say that RS\,CVn type stars are astrophysical laboratory for studying the effect of binarity on activity and vice versa.

Cool starspots typically cause brightness changes of a few tenths of a magnitude due to the rotation of the active component in an RS\,CVn system. Although the contribution of the bright faculae is only a small fraction of the total rotational brightness change (in shorter wavelength passbands), in some cases a facular network should be taken into account in order to interpret the so-called color anomaly, when the increasing surface temperature in shorter wavelengths (e.g., in $B$ passband) tends to "compensate" the color change due to cool spots \citep{1986A&A...166..167S, 1992SoPh..142..197P}.

The dynamo mechanism operating in the background, which produces starspots and other surface structures, causes changes not only on a rotational but also on a decadal time scale, just like the solar dynamo (although at a different level). \citet{2014A&A...572A..94O} investigated three overactive spotted K giants (IL\,Hya, XX\,Tri, and DM\,UMa), all of which are members of RS\,CVn binaries, whose brightness changes are also significant on a rotational and longer time scales.
Such large changes in brightness on both time scales cannot be explained simply by putting a mixture of dark spots and bright faculae on the surface. In the cited paper, the authors concluded that the anomalous brightness change on the decadal scale can be explained by a few percent change in the size of the star due to the strong and changing global magnetic field during the activity cycle.
However, except in the case of IL\,Hya, this possible explanation has not yet been confirmed.

\begin{figure*}[thb]
    \centering\includegraphics[width=1.0\textwidth,clip]{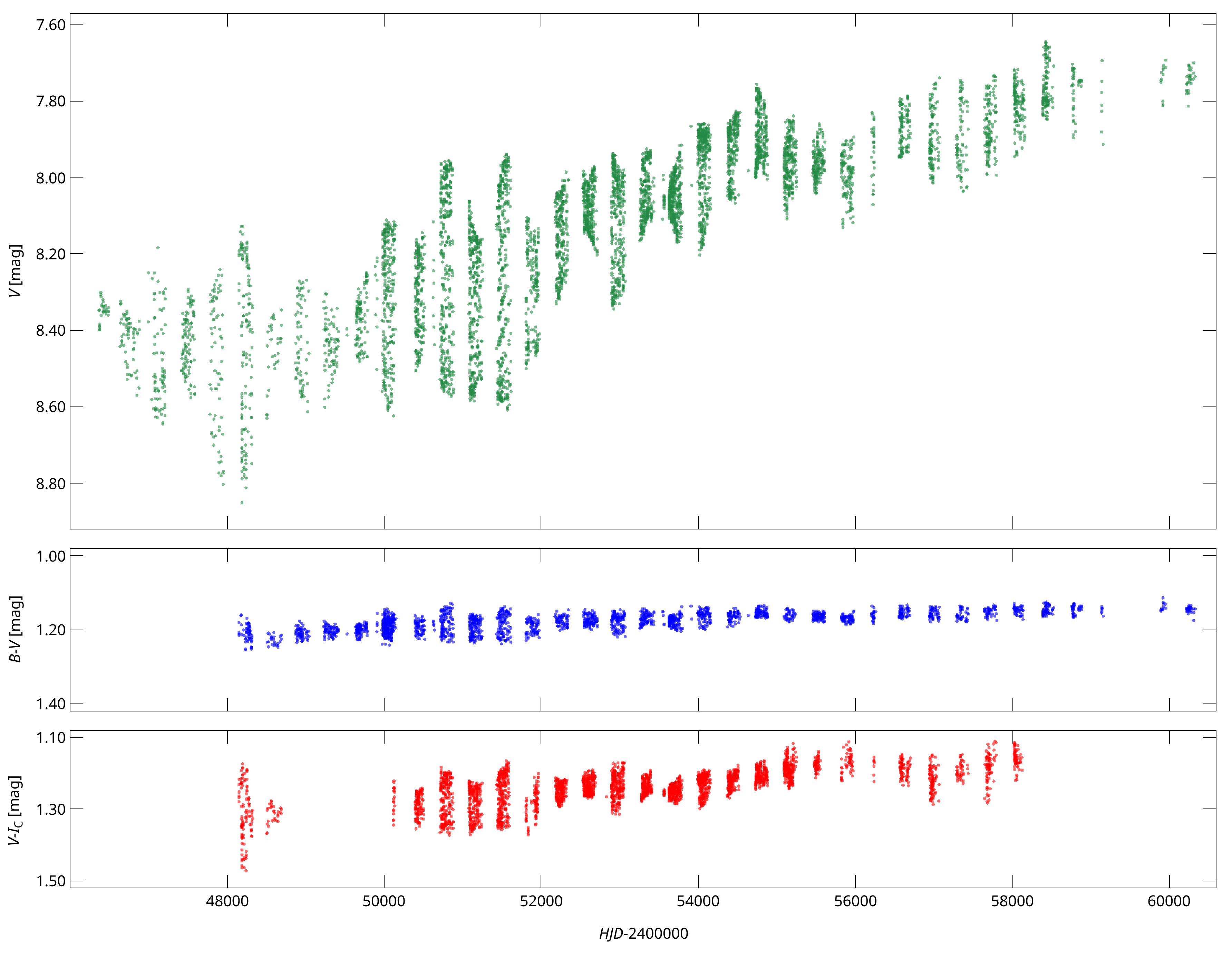}
      \caption{Combined $V$, $B-V$ and $V-I_C$ data of XX\,Tri that we use in this paper.}
      \label{fig_data}
\end{figure*}

In the most extreme case of XX\,Tri, the rotation-related photometric amplitude reached 0.6\,mag in Johnson $V$ \citep{1991IBVS.3589....1N}, which not only required extremely large and cool spots, but also the distribution of the spots on the stellar surface had to be uneven as well along the rotational phase \citep[for different approaches to light curve modeling see][]{1992A&A...259..595S,1996PASP..108...68H}.
In addition, over the past 40 years the long-term peak-to-peak brightness variation of XX\,Tri has exceeded one magnitude in $V$ (cf. Fig.\,\ref{fig_data} later in this paper).
According to this, XX\,Tri deserves the title of "the most spotted star" in the sky. Our present study is dedicated to study the long-term photometric behaviour of this particularly interesting target.

XX\,Tri (=HD\,12545), a long-period RS\,CVn system of a bright ($V_{\rm br}$=7.64\,mag) K0III active primary component. Its orbital period of $\approx$23.97 days \citep{1993AJ....106.2502B,2015A&A...578A.101K} is very close to the photometric period around 23.9$\pm$0.2 days attributed to starspots on the rotating stellar surface, that is, the system is synchronized. The star, which is otherwise a single-lined spectroscopic binary with an unseen (most probably M-type main sequence) secondary component, has long been known for its strong \cahk\ emission \citep{1985IAPPP..21...53B,1995A&AS..114..287M}. 
Spectroscopic monitoring of XX\,Tri showed its H$\alpha$ line consistently in emission \citep{1990ApJS...72..191S,1993AJ....106.2502B,1997A&AS..125..263M}, suggesting an extremely high level of chromospheric activity. Accordingly, XX\,Tri is also a bright target in ultraviolet \citep{1993AJ....106.2502B,2003A&A...408..337C} and possesses a luminous X-ray corona \citep{1997ApJ...478..358D}. The lithium abundance was found to be moderately strong, $A$(Li)$\approx$1.7, supporting that XX\,Tri is a He-core burning red giant branch (RGB) star at the evolutionary stage which can be linked to lithium enrichment \citep[see, e.g.,][etc.]{2000A&A...359..563C,2011ApJ...730L..12K,2024MNRAS.tmp..677L}. Moreover, according to \citet{2019ApJ...880..125C} lithium enrichment can even be related to binarity (but on the other hand, see also \citealt{2020A&A...639A...7J}).

Based on precise $UBV(RI)_C$ photometric measurements from 1991 \citet{1992A&A...259..595S} modeled a huge ($\Delta V\approx0.5$~mag) rotational variability caused by cool starspots. Unspotted brightness was chosen as the brightest value measured up to that time, that is, $V$=8.11\,mag \citep{1991IBVS.3589....1N}. From the 0.12-mag large amplitude $V-I$ color index, \citet{1992A&A...259..595S} derived a spot temperature of $-$1100\,K relative to the assumed unspotted surface of 4820\,K. Also, the spectral type was specified to K0III, instead of G5IV as listed earlier in \citet{1988A&AS...72..291S}. The derived spot coverage, however, was so large that the spot model essentially reached the limit where the visible stellar surface barely had any room left to account for the large rotation amplitude \emph{and} the overall long-term brightness change together. If the unspotted $V$ brightness was chosen only a few tenths of a magnitude brighter, it would not have been possible to obtain any suitable spot model for the observed light curve. We note here, that knowing the subsequent brightening of the star (see later in this paper), the assumption that there would be an "unchanged" unspotted brightness value over time is obviously not tenable.

The first Doppler image of the star was presented by \citet{1999A&A...347..225S} in the turn of 1997/98, the season characterized by high-amplitude changes in brightness. For the extreme rotational variability, the Doppler reconstruction required a gigantic high-latitude cool spot, but also bright areas in the opposite hemisphere. The star's then-record brightness and record photometric amplitude were reminiscent of the solar activity, in that our Sun is also the brightest when the degree of spot coverage is the greatest (i.e., at the maximum of the solar cycle).

In our subsequent, more extensive Doppler study \citep{2015A&A...578A.101K}, we managed to obtain 36 Doppler images based on continuous spectroscopic observations spanning six years, collected with the help of the STELLA robotic observatory at Tenerife, Spain. According to the study, the surface of XX\,Tri was covered by large, high-latitude and even polar spots and occasionally smaller equatorial spots. Furthermore, based on the time-series Doppler images it was possible to monitor the morphological changes of the spots. The measured spot decay rate provided a basis for estimating the turbulent diffusivity. However, since it was not possible to separate the opposite processes occuring together (i.e., spot decay and formation), the derived value of the turbulent diffusivity may need a revision.

In our most recent work \citep[][hereafter Paper~I]{2024NatCo..15.9986S} we presented a time series Doppler imaging study of XX\,Tri, covering 16 years (2006-2022, including the six years of data from the previous study). From this most comprehensive Doppler imaging study so far, in which 99(!) Doppler images were reconstructed from the same target, it was concluded that {\it i)} the stellar dynamo working in the background is aperiodic, likely chaotic in nature; {\it ii)} the Doppler images indicate "missing" (i.e. blocked) flux due to cool spots of up to 10\% of the total flux, which is redistributed on a global
scale, contributing to an increase in the effective temperature on a longer timescale; {\it iii)} the rotation-induced stellar photocenter variations due to starspots pose an intrinsic limitation for astrometric exoplanet catches. 
However, the Doppler study only covers less than half of the available photomertic time series, overlapping essentially with the most recent period during which XX\,Tri is at its brightest state. In comparison, at the very beginning of the systematic photometric observations, the star was fainter by about 0.65~mag on average in $V$, while its rotation amplitude was three-four times that of today. This significant difference can definitely be linked to the change in processes related to global magnetism over decades, that is, to the change in dynamo operation.
Therefore, in the present paper, we focus on the analysis of the most complete long-term photoelectric photometric datasets available from the past approximately four decades, obtained mostly with automatic photoelectric telescopes (hereafter APTs). Below we adopt from Paper~I the most thorough summary of the stellar parameters so far, which we list in Table\,\ref{tab_param}.

This paper is organized as follows: we present our available photometric data in Sect.\,\ref{sect_obs} and the methods used in Sect.\,\ref{sect_methods}. 
The seasonal photometric period changes are analyzed in Sect.\,\ref{sect_seasonalP}, long-term cycles are investigated in Sect.\,\ref{sect_tifran}, while the long-term color changes are presented and interpreted in Sect.\,\ref{sect_hrd}. Our basic results are discussed in Sect.\,\ref{sect_disc} and summarized in Sect.\,\ref{sect_sum}.

\begin{table}
 \centering
\caption{Relevant astrophysical data of XX\,Tri adopted from Paper~I.}
\label{tab_param}
\begin{tabular}{l l}
\hline\hline
\noalign{\smallskip}
Parameter & Value \\
\hline
\noalign{\smallskip}
Spectral type           & K0III \\
Distance [pc]    &  $196.3\pm1.1$ \\
$V_{\rm br}$    [mag]       & 7.64 \\
Average $(V-I_{\rm C})$     [mag]        & 1.18 \\
$P_{\rm rot}$ [d]\tablefootmark{a}   &           $23.470-24.716$ \\
$P_{\rm orb}$ [d]   &           $23.9674\pm0.0005$  \\
Inclination [\degr]            &       $60\pm10$   \\
$v\sin i$ [\kms]    &       $20.0\pm0.6$  \\
$T_{\rm eff}$ [K]\tablefootmark{b} &     4627$\pm$26 \\
$\log g$ (cgs) &      2.82$\pm$0.05 \\
Metallicity [Fe/H] &  $-$0.13$\pm$0.04  \\
Radius      [$R_{\odot}$]\tablefootmark{c}           &      $10.95^{+1.4}_{-0.95}$    \\
Radius      [$R_{\odot}$]\tablefootmark{d}           &      $8.95\pm0.23$    \\
Luminosity [${L_{\odot}}$]\tablefootmark{d}         & $337\pm2$ \\
Mass          [$M_{\odot}$]           & $1.1^{+0.2}_{-0.3}$  \\
Age [Gyr]  & $7.4^{+2.3}_{-3.2}$ \\
\noalign{\smallskip}
\hline
 \end{tabular}
\tablefoot{\\
\tablefoottext{a}{This paper, extreme values, cf. Fig.\,\ref{fig_per-ampl} and Table\,\ref{tab_periods};}\\
\tablefoottext{b}{From fitting 16 years of spectroscopic data with synthetic spectra;}\\
\tablefoottext{c}{Derived from $v\sin i$, $P_{\rm rot}$, and $i$;}\\
\tablefoottext{d}{Derived from $V_{\rm br}$, $T_{\rm eff}$, and Gaia DR3 parallax.}
}
\end{table}


\section{Observations}\label{sect_obs}

For our study, we attempted to collect all available photometric data about XX\,Tri. All published photometric data from the literature were gathered and used. Sources of the observations until 1996 are listed in \citet[][see their Table~5]{1997A&AS..125...11S}, and the corresponding data can be found in the VizieR online catalog of the Strasbourg astronomical Data Center (CDS).
Unpublished data referred as "private communication" in the above paper, including the  $UBV(RI)_C$ data from Konkoly Observatory, have been uploaded to the VizieR catalog as a supplement to this paper.

Most of the $B, V$ observations were made by the Tennessee State University T3 0.41-m Automated Photometric Telescope (APT) located at Fairborn Observatory, Arizona, USA \citep{1995ASPC...79...37H}; this series of observations ended in 2024. The $V$ observations made between 1990 October and 2015 February have been published in \citet{2017ApJ...838..122J}. 
Now all XX Tri data from T3 have been compiled and uploaded to the VizieR online catalog.

Fairborn Observatory gave place to two more APTs used for XX\,Tri photometry, the Wolfgang (T6) and Amadeus (T7) 0.75-m automatic twin telescopes, which were jointly operated initially by the University of Vienna and later by Leibniz-Astrophysical Institute Potsdam (AIP). The Wolfgang photometer is optimized for Str\"omgren $uvby$. The Amadeus photometer is optimized for the Johnson $BVR$ system. The Wolfgang system was available for XX\,Tri from 1997 to 2009. Unfortunately, however, the photometric performance of Amadeus was not satisfactory in the following ten years either. After the cathode tube was replaced with a blue-sensitive one in 13 January 2018 ($JD$\,2458132), the system performed better for a short time, but faded again shortly afterwards. The telescope finally ended its operation in January 2019. 
We note that the photometric data collected with the T6-T7 telescopes have already been published in a supplement to Paper~I\footnote{Available at \protect\url{https://data.aip.de/projects/xx_tri.html}}

The combined most complete $V$, $B-V$ and $V-I_C$ data of XX\,Tri are shown in Fig.~\ref{fig_data}. 
To convert differential measurements to absolute scale, HD\,12478 was used as a comparison star with magnitudes $V$=7$\fm$78, $B-V$=1$\fm$45, and $V-I_C$=1$\fm$38.

\section{Methods}\label{sect_methods}
\subsection{Period search by \texttt{MuFrAn}}

The period search analysis of the light curves is partially performed with the \texttt{MuFrAn} \citep[Multi-Frequency Analyzer,][]{2004ESASP.559..396C} code. This Fourier-transformation based frequency analyser tool is suitable for searching for multiple periods in unevenly sampled time series and displaying the results graphically. We use \texttt{MuFrAn} to investigate fluctuations in the rotation period over several seasons (Sect.\,\ref{sect_seasonalP}). 

Looking for the possible variability of the rotational modulation, first, the high amplitude, long-term changes were simultaneously refined with Discrete Fourier Transform \citep[DFT,][]{1975Ap&SS..36..137D} and removed from the data. The dataset prewhitened in this way was analysed again in its full length and then divided into separate subsets. We note that prewhitening is not of great importance in our case, as it has no effect in the frequency range typical of the rotation.

\subsection{Time-frequency analysis by \texttt{TiFrAn}}

To find the activity cycles and their properties the program package \texttt{TiFrAn} \citep[Time-Frequency Analysis,][]{2009A&A...501..695K} has been used. Since the amplitude of the rotational modulation shown by the observations were grossly variable, exceeding 0.6 magnitudes sometimes, we removed the rotational signals from the unevenly sampled seasonal light curves. This way we could avoid false signals in the analysis targeting decadal variability. The program package can operate with different kernels like the short-term Fourier Transform (STFT) and the Choi-Williams kernel (CWD), which has a better frequency resolution but less time resolution. A detailed description of the practical application of this tool to long-term time series of stars can be found in \citet[][see their Sect.\,3]{2009A&A...501..703O}.

\section{Results}\label{sect_res}

\subsection{Seasonal changes in the rotation period}\label{sect_seasonalP}

In the case of spotted stars, a known phenomenon is the temporal change of the rotation signal, which is mostly explained by the changing spot configuration from season to season, as well as differential rotation. On the one hand, it would be desirable to examine as short time intervals as possible (e.g., annual), but on the other hand, deriving a precise period is only possible in the case of a sufficiently long data series.
Since during an annual observing season our data cover roughly 10-12 rotation periods at most, period search for each year is not possible. Instead, we divided the data into four segments of similar length, keeping in mind that no significant changes occur within each segment, in the sense that the trend or "dynamics" of change are roughly "constant" within a segment. With the division according to this principle, the following time intervals were created ($HJD-2400000$): 46370-50152 (nearly constant average brightness), 50391-53804 (continuous increase in average brightness), 53995-56244 (average brightness reaches its local maximum and then decreases), and 56555-60327 (slight upward trend in average brightness); cf. the upper panel of Fig.\,\ref{fig_data}. 

\begin{figure}[t]
    \includegraphics[width=\columnwidth]{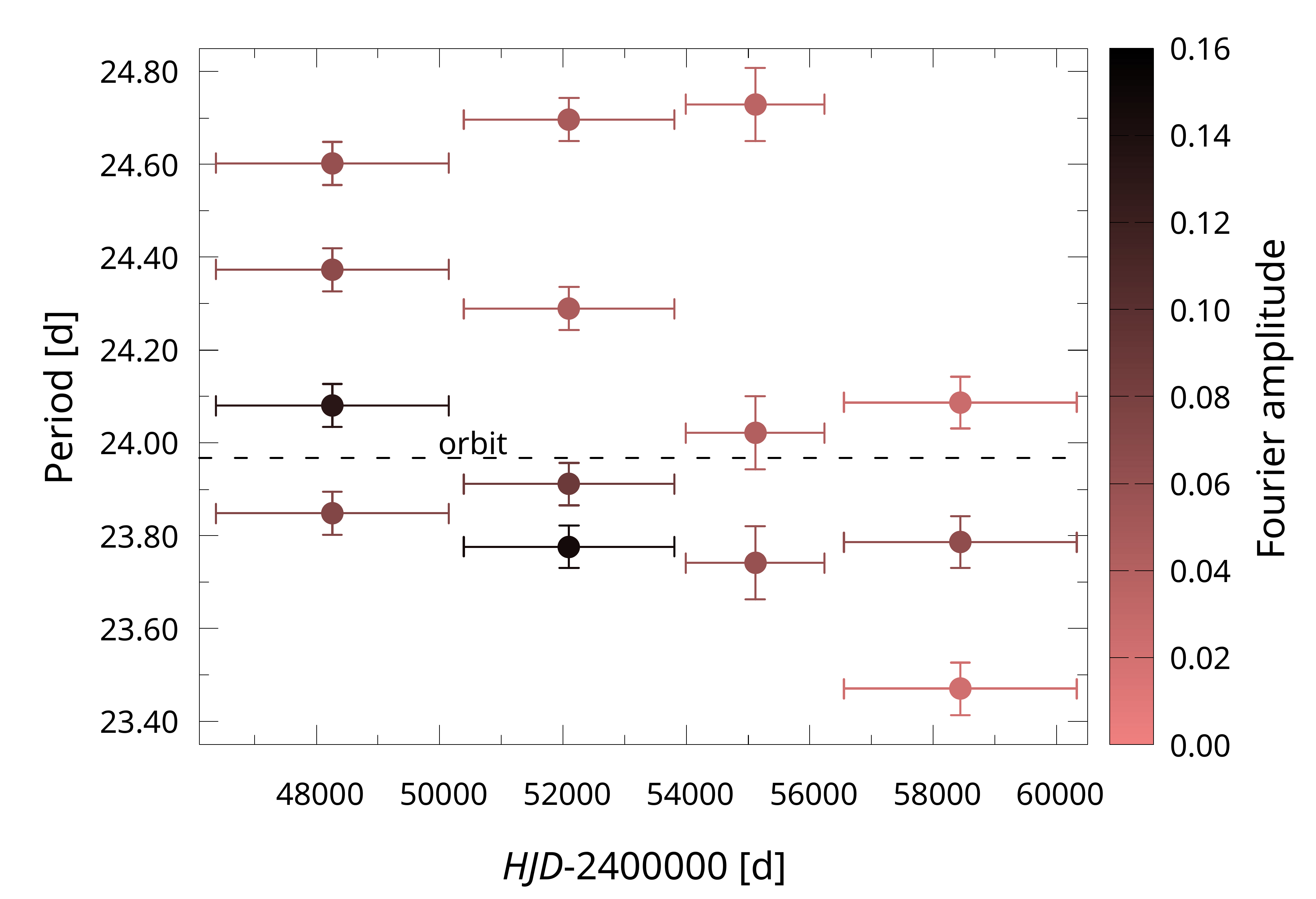}
      \caption{Rotational periods of XX\,Tri for four segments of the $V$ dataset. The orbital period is indicated with a dashed line. The dots marking the detected rotational signals of each segments are colored according to their Fourier amplitudes. See also Table\,\ref{tab_periods}.}
      \label{fig_per-ampl}
\end{figure}

The period search was carried out as follows. Using \texttt{Mufran}, we searched for the most dominant rotation-related period from data previously cleaned of long-term trends. The next period search was always performed on the spectrum prewhitened by the previously found dominant period, and so on until no period of reasonable amplitude remains. The resulting 3-4 periods per season with their respective Fourier amplitudes are listed in Table\,\ref{tab_periods} for all four intervals, which are graphically represented in Fig.\,\ref{fig_per-ampl} in comparison with the orbital period.

We note that in the classical approach, according to the uncertainty principle, $\Delta w\Delta t\geq$1 relation applies between the $\Delta w$ width of the Fourier signal and the $\Delta t$ length of the dataset. Based on this, the uncertainty of the Fourier signals obtained for the different seasons (frequency converted to days) would be on the order of $\sim$0.15-0.20\,d.
However, this value overestimates the more realistic uncertainties that can be derived from the peak widths measured at 90\% heights of the peaks in the amplitude spectra \citep[cf.][]{2003A&A...410..685O}. In comparison with the obtained periods and their associated errors, we also present the Lomb-Scargle periodograms obtained for the entire $V$ data and the four sub-seasons in Appendix\,\ref{LS_periodograms}.
The most dominant periods in Fig.\,\ref{fig_per-ampl} scatter around the orbital period, while those with smaller amplitudes usually diverge more. Possible reasons for this are discussed in Sect.\,\ref{sect_disc}.

\begin{table*}[h!!!]
\centering
\caption{Seasonal rotational periods and their corresponding Fourier amplitudes.}
\label{tab_periods}
\begin{tabular}{c c | c c | c c | c c }
\hline
\hline\noalign{\smallskip}
\multicolumn{2}{c|}{46370-50152\tablefootmark{a}} & \multicolumn{2}{c|}{50391-53804\tablefootmark{a}} &  \multicolumn{2}{c|}{53995-56244\tablefootmark{a}}  &  \multicolumn{2}{c}{56555-60327\tablefootmark{a}}\\
period [d]\tablefootmark{b}  &  amplitude   &  period [d]\tablefootmark{b}   &   amplitude  &  period [d]\tablefootmark{c}  &  amplitude &  period [d]\tablefootmark{d}  &  amplitude\\
\noalign{\smallskip} 
\hline\noalign{\smallskip} 
24.080 & 0.13268 & 23.776 & 0.14665 & 23.741 & 0.05881 & 23.786 & 0.06398 \\
23.848 & 0.07256 & 23.911 & 0.08768 & 24.021 & 0.04071 & 24.087 & 0.02494 \\
24.373 & 0.06623 & 24.696 & 0.04672 & 24.729 & 0.03543 & 23.470 & 0.02083 \\
24.602 & 0.06005 & 24.289 & 0.04480 & --     & --      & --     & --      \\
\hline\noalign{\smallskip}
\end{tabular}
\tablefoot{
\tablefoottext{a}{$HJD-$2400000;}
\tablefoottext{b}{Accuracy limit $\pm$0.046\,d;}
\tablefoottext{c}{Accuracy limit $\pm$0.079\,d;}
\tablefoottext{d}{Accuracy limit $\pm$0.056\,d.}
}
\end{table*}


\subsection{Long-term brightness changes}\label{sect_tifran}

\begin{figure*}[h!!!]
    \includegraphics[height=10.4cm]{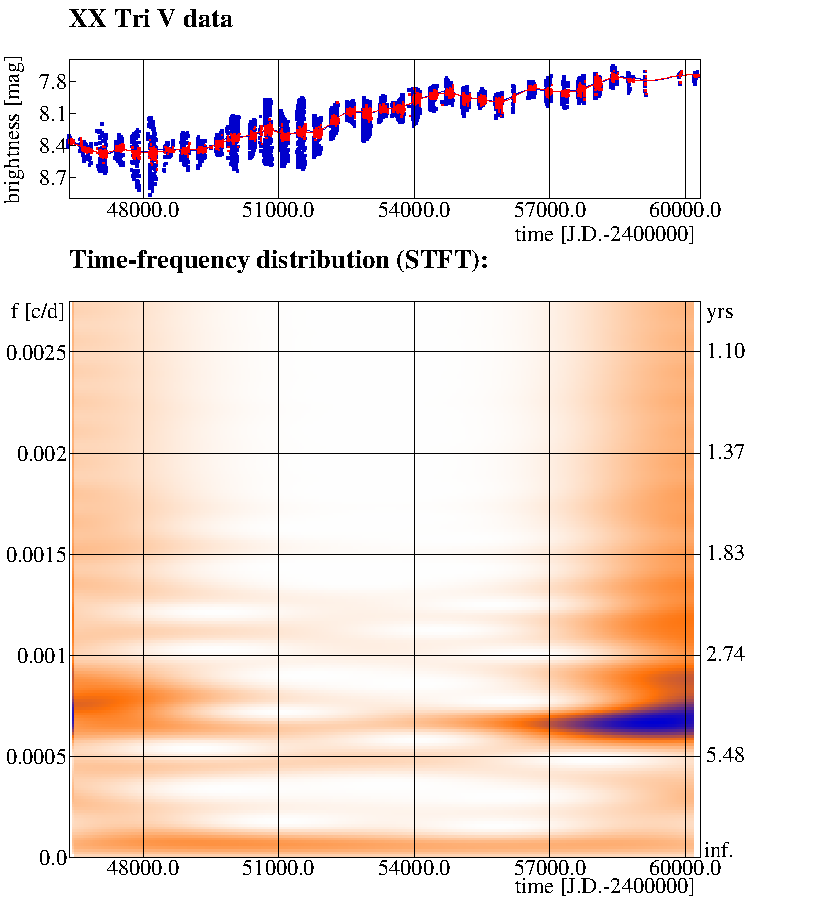}\includegraphics[height=10.4cm]{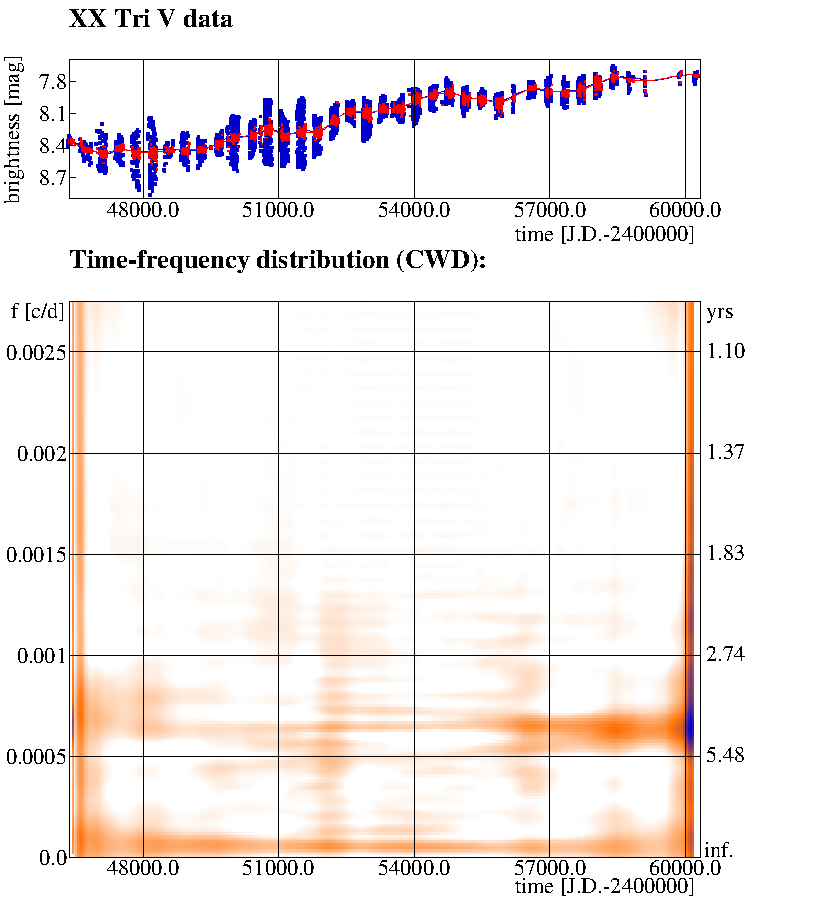}
      \caption{Time-frequency plot of the long-term variability of XX\,Tri over four decades using STFT (lower left panel) and CWD (lower right panel) kernels. In the two upper (in this case identical) panels, the blue dots show the original $V$ magnitudes, and the blue line shows their spline fit, while the red dots represent the data cleaned of the rotational signal, and the red line represents the corresponding spline fit.}
      \label{V_STFT_CWD}
\end{figure*}

\begin{figure*}[thb]
    \includegraphics[height=10.4cm]{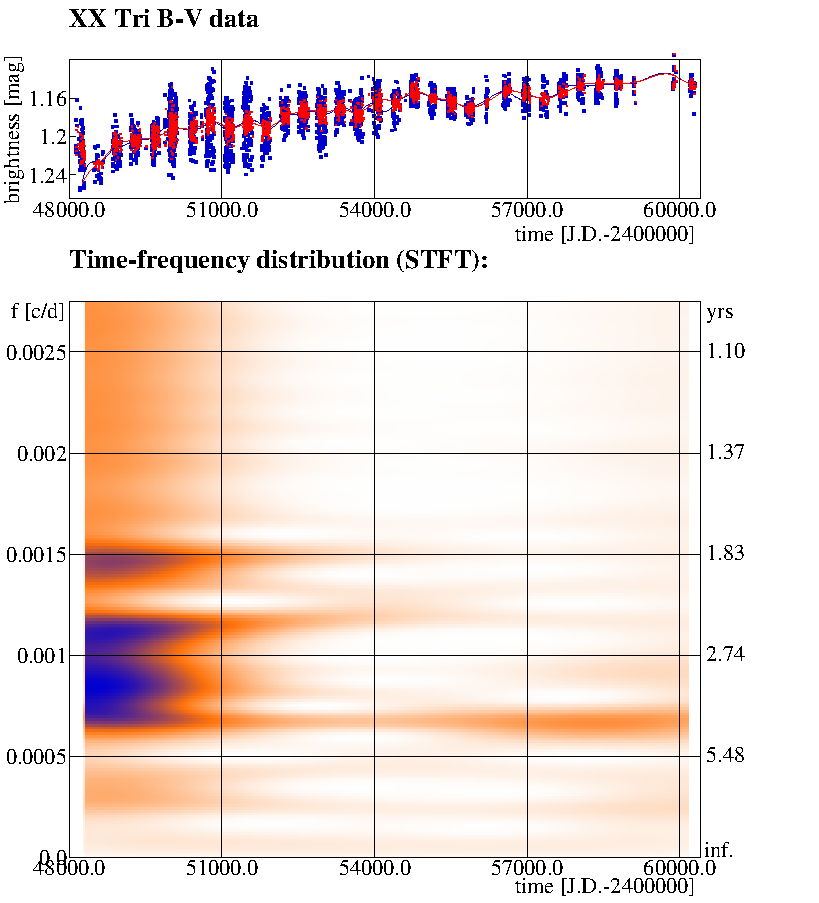}\includegraphics[height=10.4cm]{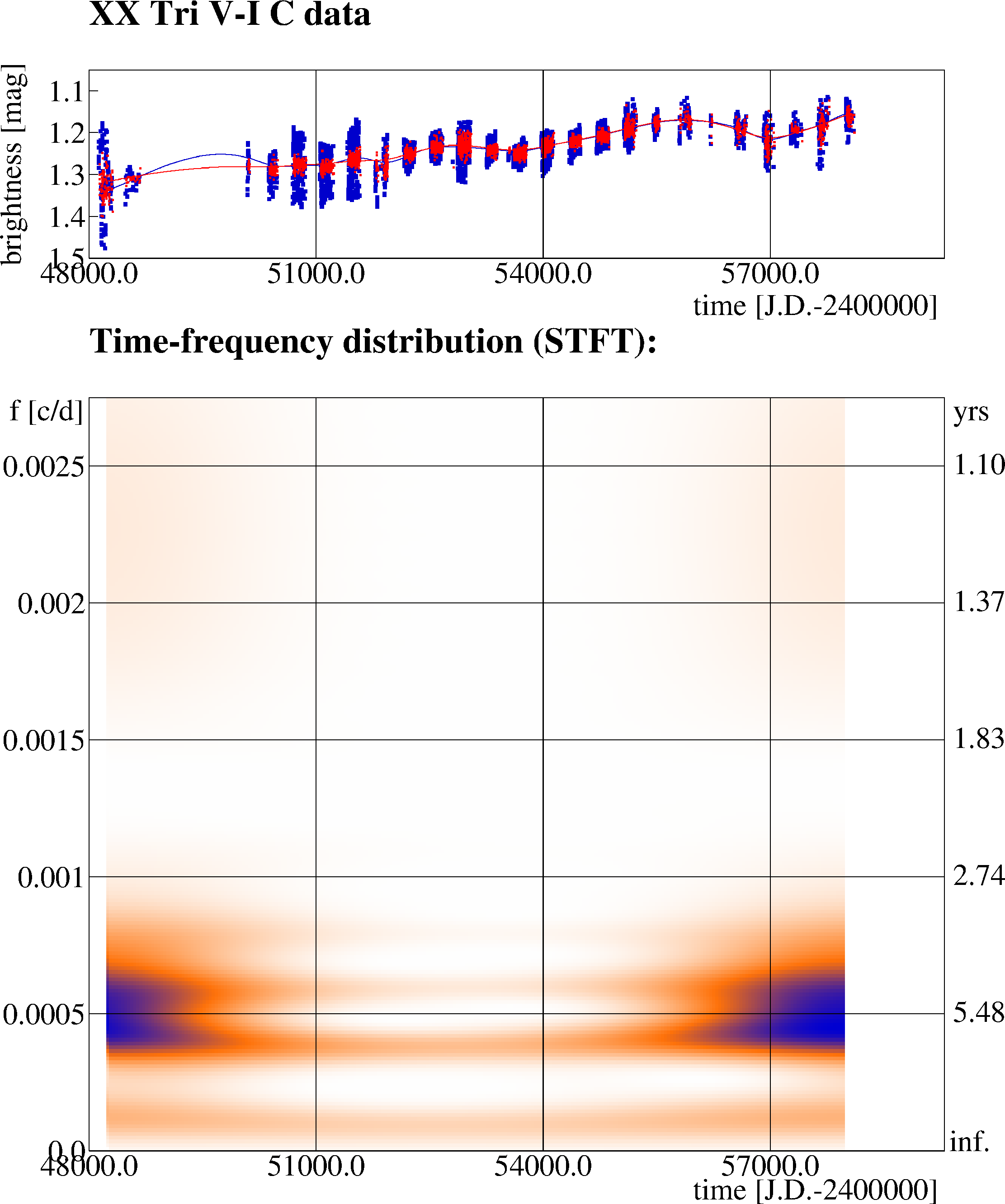}
      \caption{Time-frequency plot of the long-term variability of XX\,Tri over nearly four decades in $B-V$ (left panels) and $V-I_C$ color indices (right panels). Otherwise as in Fig.\,\ref{V_STFT_CWD}.}
      \label{BV_VI_STFT}
\end{figure*}

Long-term, cycle-like brightness variation with a characteristic time-scale of about 4 years of XX\,Tri has been derived for a 21 years long dataset by \citet{2009A&A...501..703O} using \texttt{TiFrAn}. The present photometric dataset spans now four decades for which we repeated the time-frequency analysis. First, the rotational modulation was removed from the data since the uneven distribution of the measured datapoints and the sometimes very high amplitude could alter the seasonal variations giving rise of possible false long-term signals. The result of the long-term variability with STFT is plotted in the left panel of Fig.~\ref{V_STFT_CWD}, while using CWD with better frequency but poorer time resolution, is seen in the right panel of Fig.~\ref{V_STFT_CWD}. 

Apart from the visible, very high amplitude brightening during the time-span of $\sim$40 years, a modulation in the order of $\approx$11 years, a slowly decreasing cycle between $\approx$5.7-5.2 years and a constant cycle of about 4 years are present in the data throughout the observations, the latter being the strongest signal of all.


\subsection{Migration on the HRD}\label{sect_hrd}

\begin{figure*}[h!!!]
   \vspace{1cm} \includegraphics[height=9.6cm]{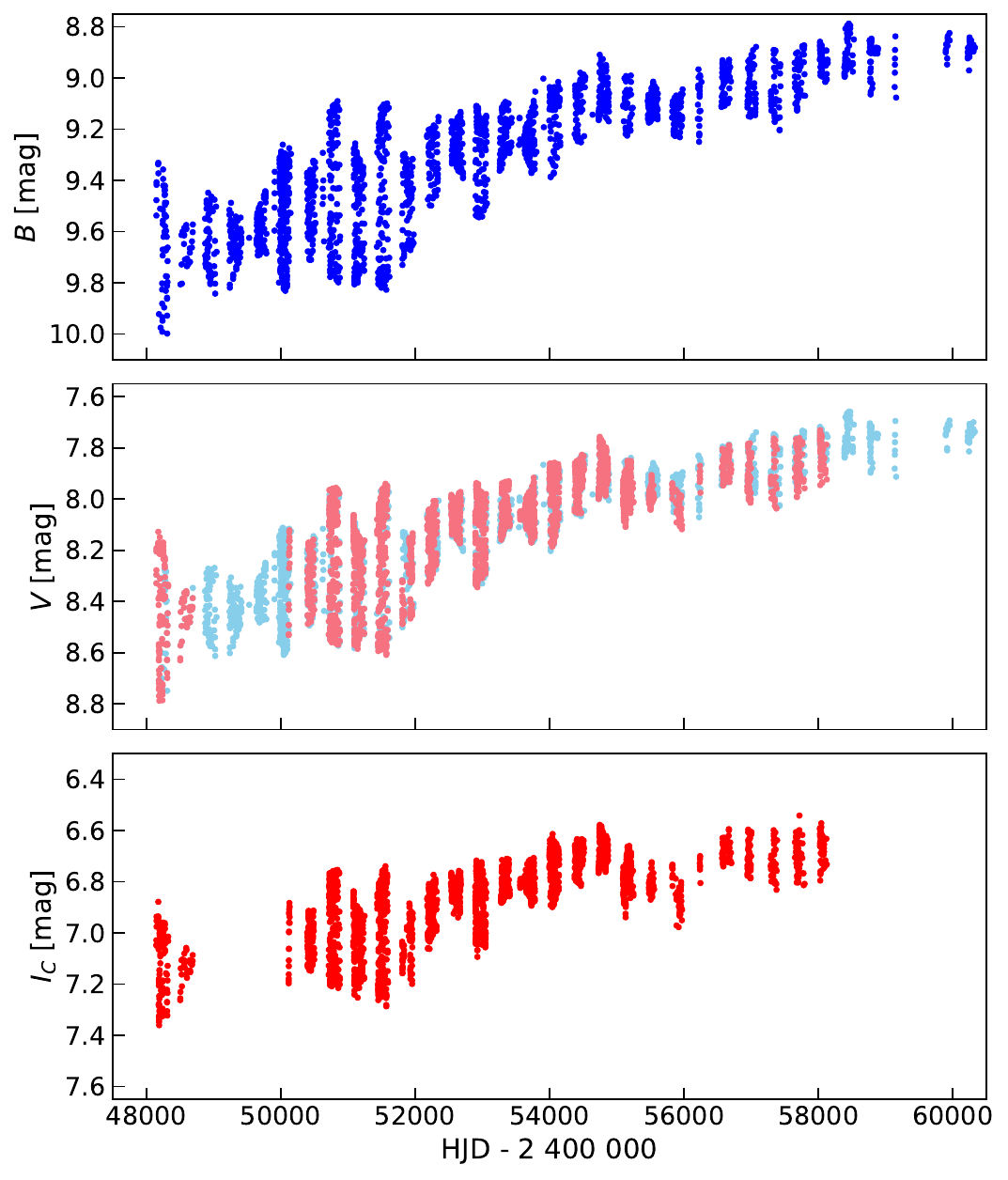}\includegraphics[height=9.6cm]{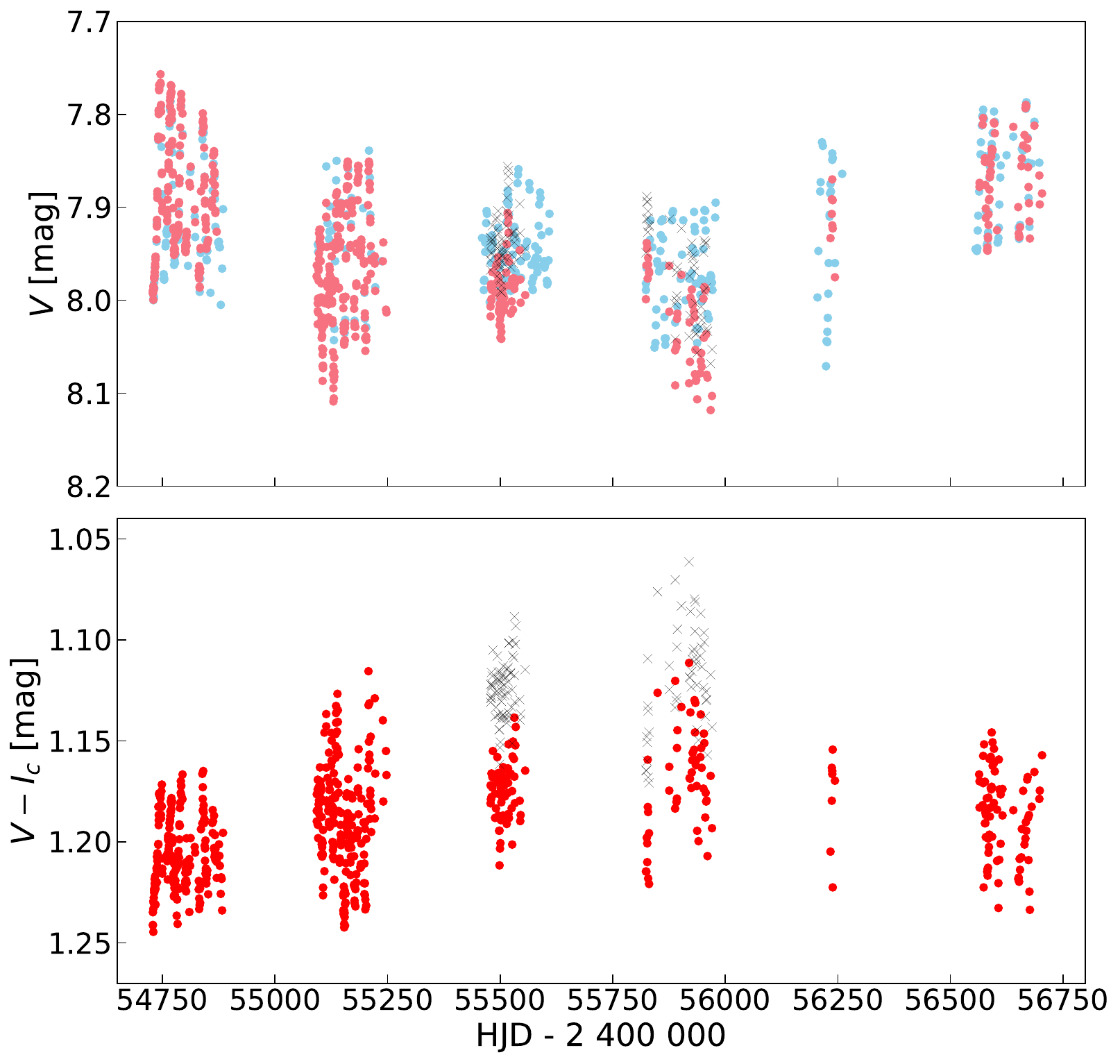}
      \caption{Left, from top to bottom: $B$, $V$ and $I_C$ data used for temperature determinations. In the middle panel showing $V$ data, points measured simultaneously with $B$ measurements are marked in light blue, and points measured simultaneously with $I_C$ measurements are marked in light red. Right: enlarged view around the discrepant $V$ (upper panel) and $V-I_C$ (lower panel) data; data points corrected by 0$\fm$05 are indicated by black crosses (for details see the first paragraph of Sect.\,\ref{sect_hrd}).}
      \label{temp_BVI_colors}
\end{figure*}

Long-term $BVI_C$ dataset is used to trace the temperature changes on the surface of XX\,Tri. 
The bulk of the dataset consists of independent series in $B, V$ and $V, I_C$ colors by different automated telescopes (see Sect.\,\ref{sect_obs}), supplemented with a small amount of data gathered from other sources. As Fig.~\ref{temp_BVI_colors} shows, the agreement between the two sets of $V$ measurements made by different telescopes is excellent, apart from an about 0$\fm$05 discrepancy between $HJD$\,2455478 and 2455970, for unknown reason to us. This, however, translates to about 80\,K temperature difference which is not negligible. The right panel of Fig.\,\ref{temp_BVI_colors} shows the shifted $V$ data and the corresponding shifted $V-I_C$ values (all of the original color index data are shown in Fig.\,\ref{fig_data}).

The color index data $(B-V, V-I_C)$ were corrected for interstellar extinction to define temperatures. For getting the extinction parameter we accessed the 3D dust maps of interstellar dust reddening\footnote{\protect\url{http://argonaut.skymaps.info/}} \citep[Bayestar17,][]{2018MNRAS.478..651G}. According to the best-fit distance-reddening curve we adopt a color excess of $E(B-V)=0.05^{+0.02}_{-0.03}$ \citep[cf.][see also their Fig.\,4]{2014A&A...572A..94O}, along with $E(V-I_C)/E(B-V)=1.25$ from \citet{1999PASP..111...63F}.

\begin{figure}[thb]
    \includegraphics[width=\columnwidth]{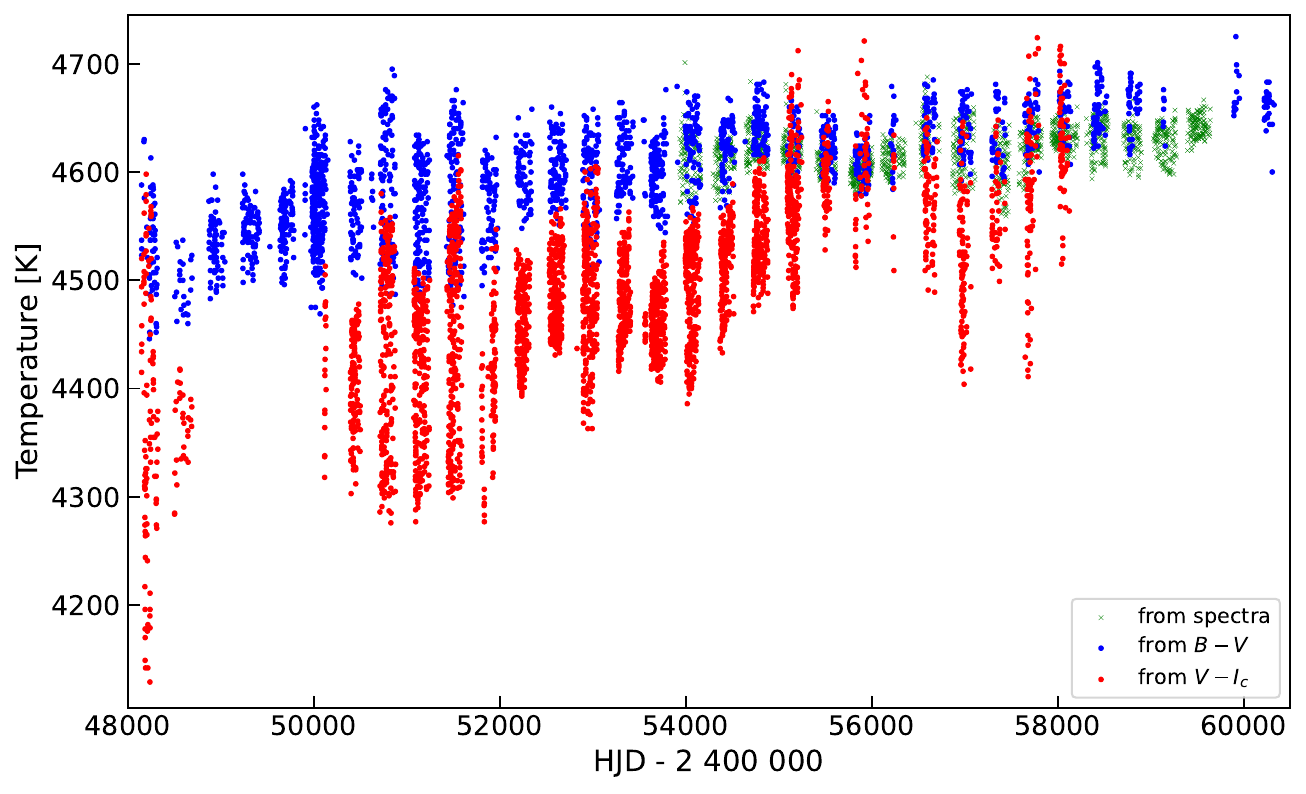}
      \caption{Comparison of the overall stellar temperature variations derived from the STELLA spectra used in Paper~I (green crosses) and from $B-V$ (blue dots) and $V-I_C$ color indices (red dots).}
      \label{temp_color}
\end{figure}

Surface temperatures were estimated using the empirical color-temperature calibration of \citet{Worthey2011ApJS..193....1W} by interpolating their table to the the measured color index values corrected for the interstellar extinction and the surface gravity ($\log g = 2.82$) and metallicity ([Fe/H]=$-0.13$). For the second half of the available color-temperature data we have surface temperature values obtained from spectral synthesis of 1822 spectra (Paper~I). The results are plotted in Fig.\,\ref{temp_color}.

\begin{figure}[t!!!]
    \includegraphics[width=\columnwidth]{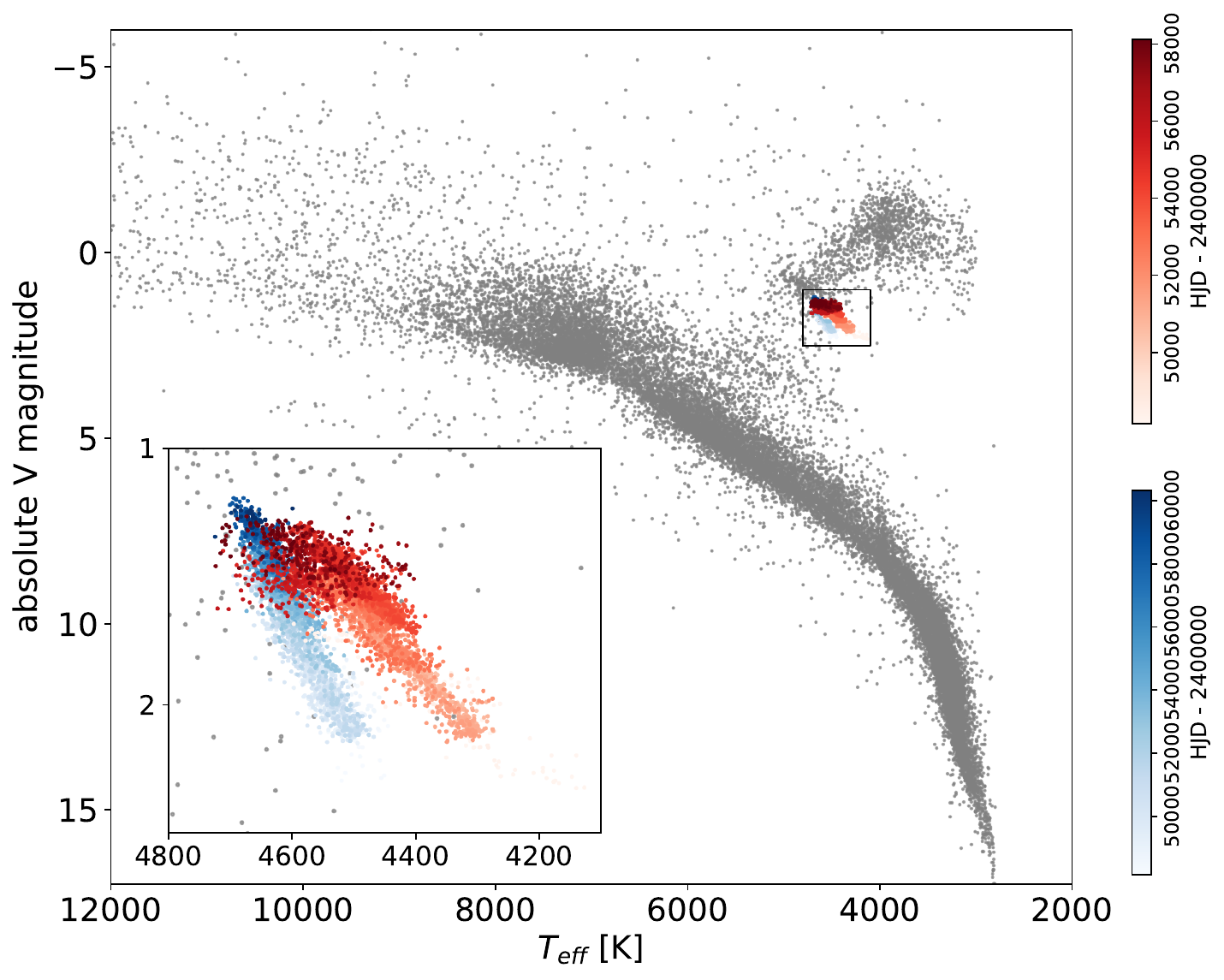}
      \caption{Change of the location of XX\,Tri on the HRD due to temperature variation derived from $B-V$ (bluish shades) and $V-I_C$ (reddish shades). The passage of time is indicated by the color scales on the side.}
      \label{temp_HRD}
\end{figure}

The substantial variability of the surface temperature in the order of about 500\,K seen in Fig.~\ref{temp_color}, apart from the interstellar extinction, affect the position of XX\,Tri in the HRD as was already shown in \cite{2014A&A...572A..94O}. Here we present stellar surface temperatures derived from $B-V$ and $V-I_C$ color indices. The paths of the star on the HRD in the course of the decades long variation according to the two series of temperature values are plotted in Fig.~\ref{temp_HRD}.

\begin{figure}[h!!!]
    \includegraphics[height=10.4cm]{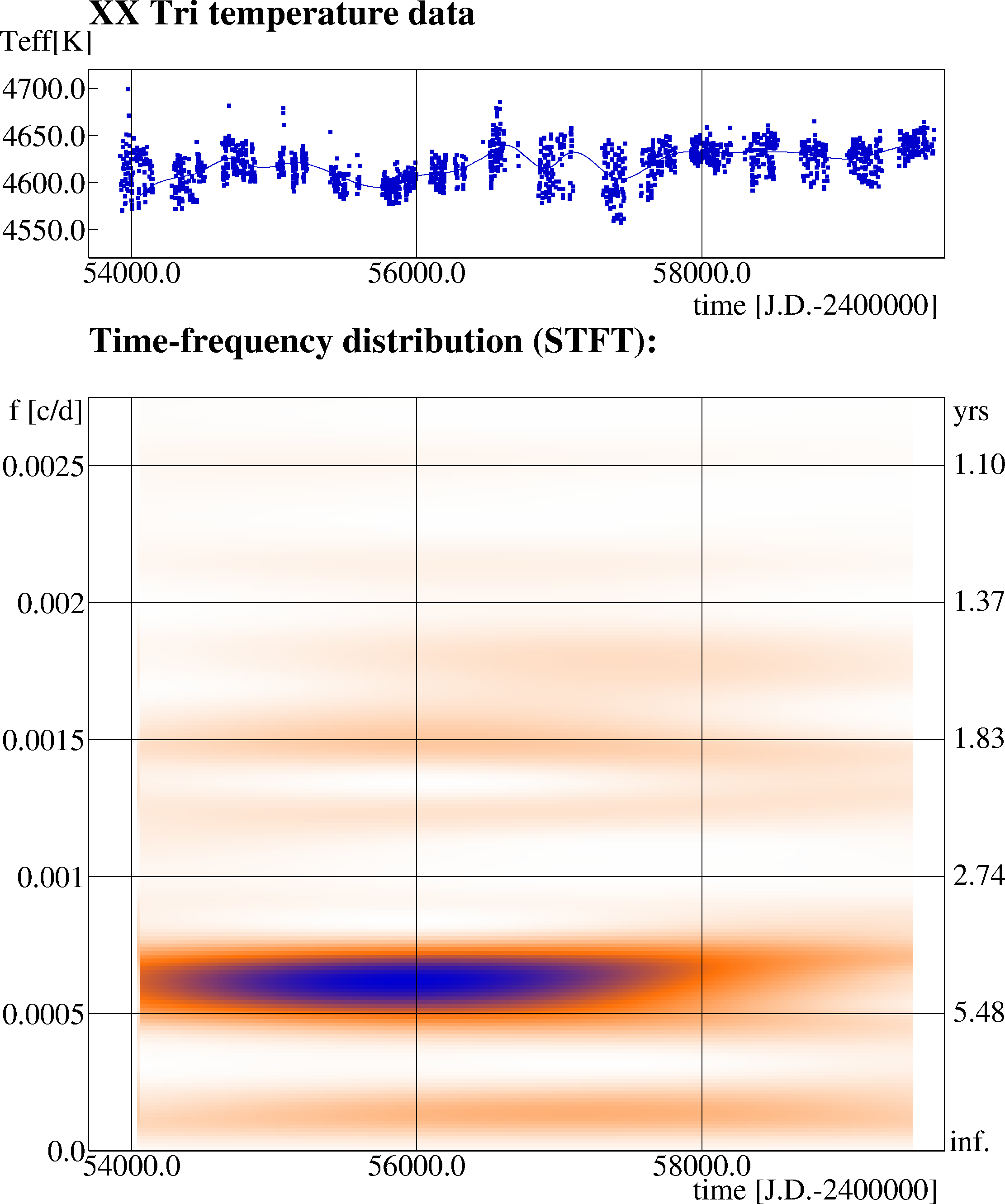}
      \caption{Variability over 16 years of the spectroscopically determined effective temperature of XX\,Tri taken from Paper~I. Otherwise as in Fig.\,\ref{V_STFT_CWD}.}
      \label{longterm_temp}
\end{figure}

\section{Discussions}\label{sect_disc}

\subsection{Long-term periods}

It is widely known that in (otherwise usually tidally locked) RS\,CVn systems, activity cycles are typically anharmonic \citep[e.g.][see also the references therein]{2010IAUS..264..120L}, and in many cases, multiple and variable quasi-cycles can be found \citep{2009A&A...501..703O}. The photometric data of XX\,Tri perfectly support this observation: we see a main $\sim$4-year cycle that varies slightly over time, as well as a $\sim$6.5-year cycle that appears spectacularly in the $V-I_C$ color index and although weaker, it is also present in the $B-V$ data (see Fig.\,\ref{BV_VI_STFT}). In the latter (longer) data series, a $\sim$11-year cycle, the double of 6.5 years, can also be detected. We note that the only cycle-like variation found earlier by \citet{2009A&A...501..703O} in their 21 year-long photometric dataset was 3.8 years long, very close to our $\sim$4-year main cycle. We also note that, even though the $B-V$ and $V-I_C$ color index data are 5 and 12 years shorter, respectively, than the full set of $V$ observations, the cycle timescales and the time-frequency patterns revealed are very similar to those found in the $V$ data (cf. Figs.\,\ref{V_STFT_CWD}-\ref{BV_VI_STFT}).

In our recent spectroscopic analysis covering 16 years (Paper~I) we found a cycle period of 1514$\pm$83 days ($\approx$4.1 yr) in the effective temperature change.
The coincidence of the $\sim$4-year cycle found in the long-term brightness with this 4.1-year period is striking, which is also supported by the agreement with the cycles of similar lengths found in the $B-V$ and $V-I$ color indices, respectively; see Figs.\,\ref{temp_color}-\ref{longterm_temp}. 
This mid-term cycle of $\sim$4 years does not appear to be closely affected by the longer-term observed trends as seen in the $V$ data. In other words, this $\sim$4-year cycle does not show any significant change with the total spot coverage, one only needs to take a look at the huge amplitude brightness changes characteristic of the first half of the photometric observations.
In this regard, we suspect that this cycle is likely the result of a flip-flop-like phenomenon, that is, a longitudinal rearrangement of large spots rather than a large-scale trend reversal in total spottedness. 

\begin{figure*}[thb!!!]
    \includegraphics[width=2\columnwidth]{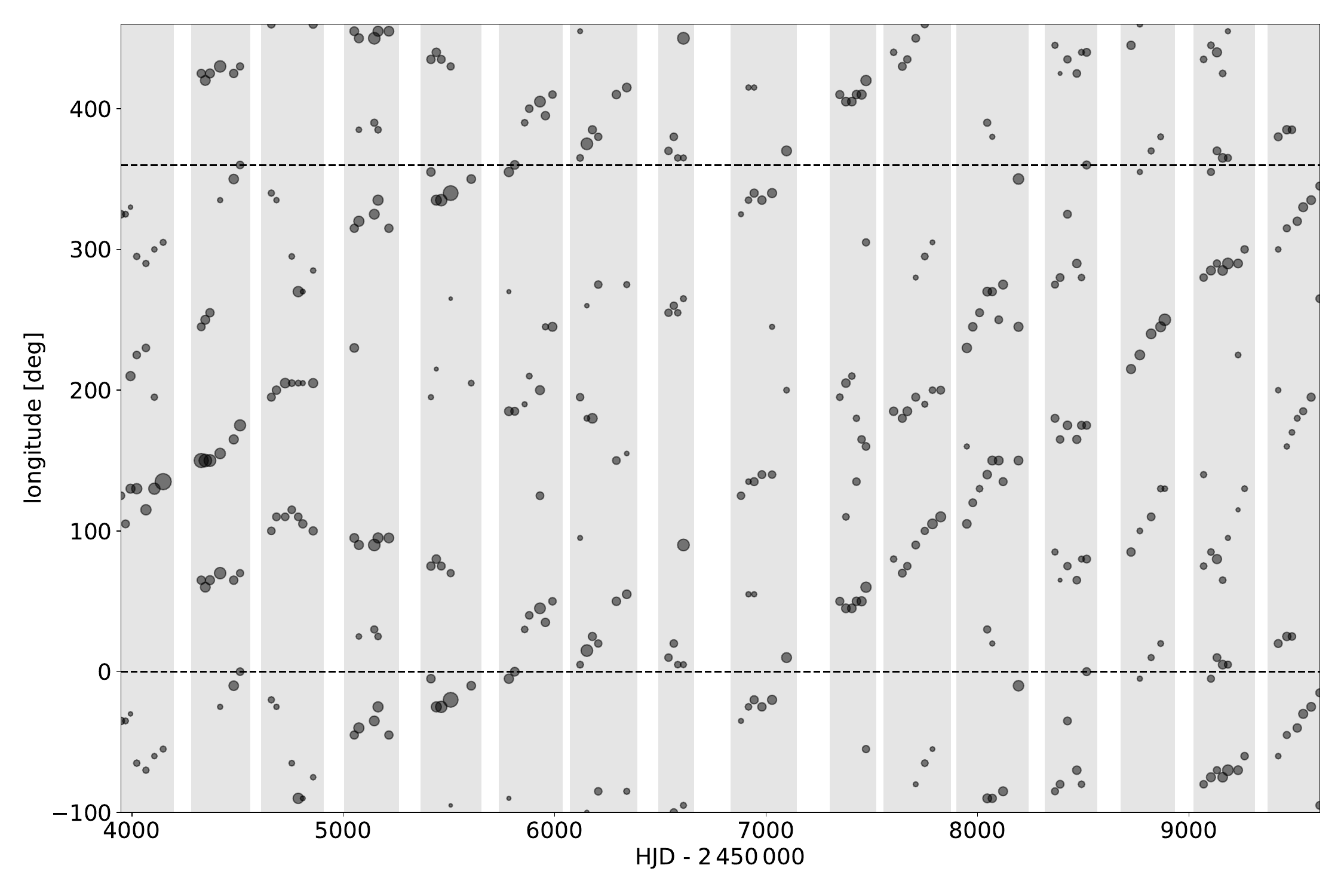}
      \caption{Longitudes of the activity centers derived from 16 years of time-series Doppler images. We identified two or three activity centers per image. The larger the dot size, the more dominant the active longitude. Vertical gray bars indicate the separate observing seasons.}
      \label{activelongitudes}
\end{figure*}

Based on the most comprehensive Doppler imaging study ever applied to a spotted star, also covering the longest period of time so far, that is, the 16 year-long time-series Doppler imagery of XX\,Tri in Paper~I \citep[see Fig.\,2 and the supplementary movies in][]{2024NatCo..15.9986S}, it appears that typically 2-3 activity centers (active longitudes) are present at any given time, the positions of which change slowly but continuously. That is, they are not closely tied to the orbital phase.
To examine this in more detail, we performed the following study. Based on the time-series Doppler images, we calculated the longitude values around which the spots were grouped ("active longitudes"). For this, we prepared a distribution function of the spot filling factor $f$ along the longitude scale for each temperature map \citep[for details on how to calculate $f$ see Eq.\,5 in][]{2024A&A...684A..94K}. The longitudinal values of the first and second (and if relevant, the third) maxima of the histograms were considered active longitudes. The changes of these active longitude values over 16 years are shown in Fig.\,\ref{activelongitudes}. In the figure, active longitudes appear as separate strings of dots, which can be followed for three to eight seasons, depending on their lifespan. A relatively constant drift is noticeable (especially from the second half of the 16 years covered, with a slope of about 0.12$^{\circ}$day$^{-1}$), which is due to the difference between the dominant photometric period and the orbital period and from which it appears that there are no activity centers bound to the orbit. It is also clear that the structures characteristic of the period between the fourth and eighth seasons differ significantly from those of the periods before and after it. The  average lifetime of the active longitudes together with the characteristic timescale of the global changes reflected in Fig.\,\ref{activelongitudes} support that the roughly 4-year mid-term cycle may indeed be related to significant rearrangements of the spots, that is, a flip-flop-like behavior.

The above, however, suggests a more complex behavior than simply a swapping of two activity centers located in opposite hemispheres. It is very likely, that the underlying dynamo, best represented by a chaotic oscillator \citep[cf.][etc.]{2009SSRv..144...25S,2014SSRv..186..525A} has complex nonlinear feedback mechanisms, although their characteristics are not precisely known, not only for stellar dynamos, but even for the solar dynamo. With all this, the main conclusion of Paper~I that the long-term surface activity of XX\,Tri is mostly chaotic in nature can still be maintained.

\subsection{Seasonal period changes and differential rotation}

Based on the period search presented in Sect.\,\ref{sect_seasonalP}, it becomes clear that some slightly different, but certainly rotationally induced, periods are simultaneously present. The most obvious explanation for this phenomenon is that the star rotates differentially, while the starspots from which the rotation signals come are located at different astrographic latitudes. Accordingly, the $\Delta P/\overline{P}$ value calculated from the $\Delta P$ interval of the different rotation periods and their $\overline{P}$ average gives a reasonable estimate to the surface shear parameter $\alpha=(\Omega_{\rm eq}-\Omega_{\rm pol})/\Omega_{\rm eq}$, calculated from the equatorial (eq) and poloidal (pol) angular velocities, characterizing the differential rotation. Using the values in our Table\,\ref{tab_periods} yields $\Delta P/\overline{P}$$\approx$0.05, supporting the general picture emerging from previous experiences \citep[e.g.,][etc.]{2004MNRAS.348.1175P,2007AN....328.1047M,2007AN....328.1075W,2016A&A...593A.123O,2024ApJ...976..217X} that the surface differential rotation for active giants in tidally bound RS\,CVn systems is typically larger than that seen in rapidly rotating dwarfs, but does not reach either that of single giants with similar rotation rates or the value measured for the Sun \citep[see the comprehensive study by][]{2017AN....338..903K}.

If we compare the values of Table\,\ref{tab_periods} with the overall brightness changes presented in Fig.\,\ref{fig_data}, it is seen that, moving forward in time, in the second half of the observations, smaller amplitude changes associated with higher average brightness typically have shorter rotation periods, cf. Fig.\,\ref{fig_per-ampl}. This can be interpreted as follows. At medium inclination ($\sim$60$^{\circ}$ like in the case of XX\,Tri), spots closer to the visible pole result in greater spot coverage (greater average brightness decrease) but also a smaller brightness variation amplitude, while at lower latitudes, spots may even disappear at some phases during rotation, so they mainly increase the amplitude and contribute only slightly to the overall brightness decrease.
That is, in the first half of the  entire observation period of four decades, the spots were present both at lower latitudes and near or even covering the pole, while in the second half, spots did not cover the pole, rather appeared at mid-latitudes. This can be brought into line with the assumption of solar-like differential rotation working on the surface of XX\,Tri, that is, shorter period signals originate from lower latitudes, while longer ones come from higher latitudes. This assumption is supported by the results of an earlier Doppler imaging study and the recent one in Paper~I. During the first half of the photometric observations, more precisely during the high-amplitude period around $\sim$$HJD$\,2450820 (i.e., 1998), the only reconstructed Doppler image revealed a huge asymmetric polar cap \citep{1999A&A...347..225S}, while the time-series Doppler imaging study in Paper~I, in which the spectroscopic data essentially overlap with the second half of the photometric data, shows that the centroids of the spots are typically located around latitudes between 45$^{\circ}$ and 65$^{\circ}$, but the spots do not extend so far as to cover the visible pole.

\begin{figure}[thb!!!]
    \includegraphics[width=\columnwidth]{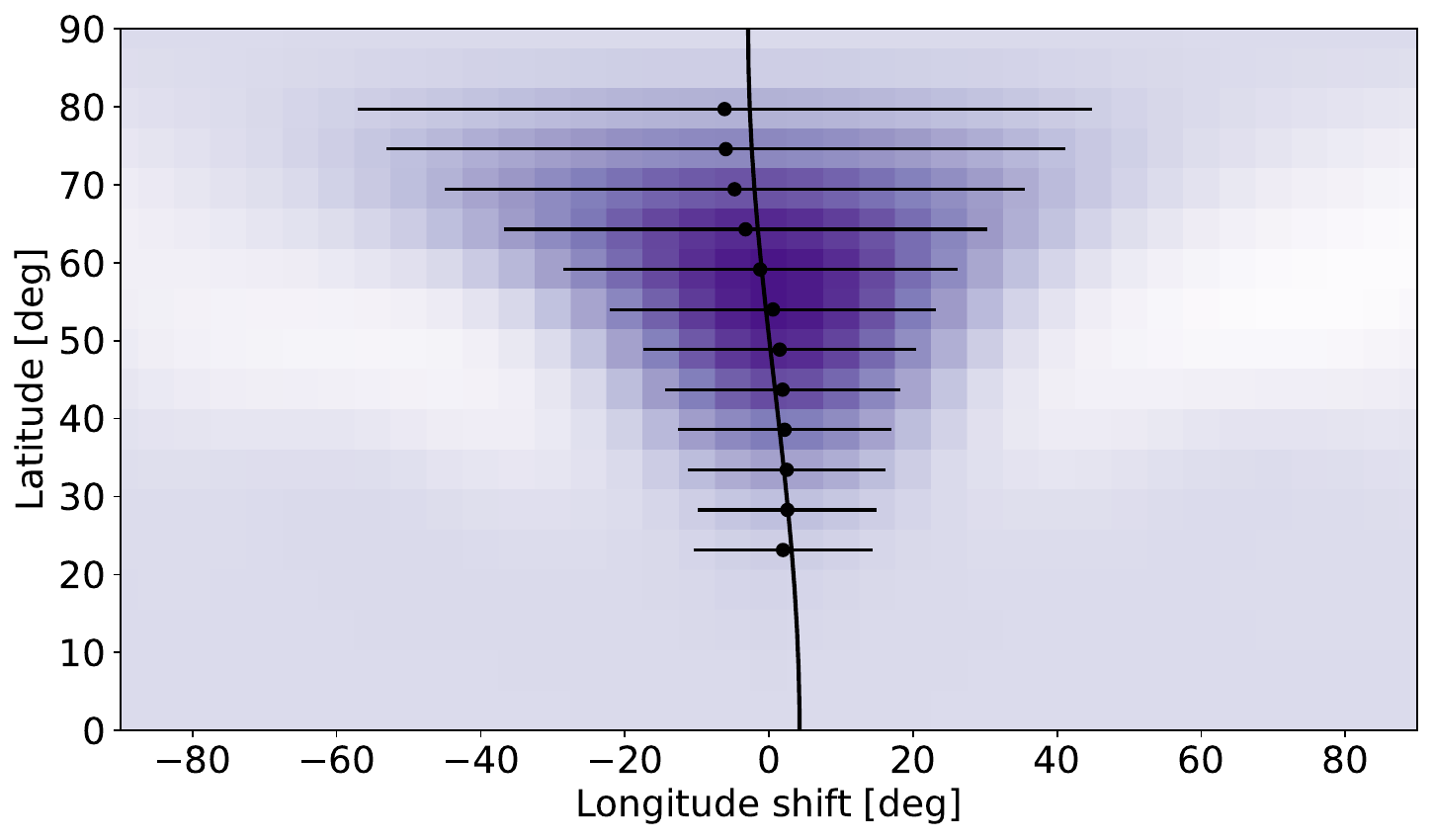}
      \caption{Grand average of the cross-correlation function maps obtained from \texttt{ACCORD}. The map indicates how much longitude shift occurs at a given latitude due to surface shear over 35 days (the median time interval between successive Doppler maps). The solid line is a sun-like differential rotation fit with a shear coefficient of $\alpha=0.014$.}
      \label{ccf}
\end{figure}

To further confirm the operation of solar-type surface differential rotation on XX\,Tri,  using the time-series Doppler images from Paper~I, we applied the method of average cross-correlations, a well-established technique dubbed \texttt{ACCORD} \citep[e.g.,][]{2012A&A...539A..50K,2015A&A...573A..98K,2017A&A...606A..42K} for detecting surface shear on spotted stars. \texttt{ACCORD} combines information from spot displacements on successive Doppler images to detect and measure surface differential rotation. As shown in Fig.\,\ref{ccf}, the method results in a grand average correlation pattern clearly indicating a weak sun-like shear. Fitting the pattern by a rotation law in the usual form $\Omega(\beta)=\Omega_{\rm eq}(1-\alpha\sin^2\beta)$, where $\Omega(\beta)$ is the angular velocity measured at latitude $\beta$, yields a value of 0.014$\pm$0.003 for the shear coefficient $\alpha$.
We draw attention to the fact that the $\alpha$ value obtained from the time-series Doppler images, although smaller, does not contradict the $\Delta P/\overline{P}$$\approx$0.05 value given above based purely on photometry, as the latter is only a rough estimate of the relative shear. Finally, we note that in our upcoming paper we are preparing to present a more detailed study focusing on the meanwhile expanded number of time-series Doppler images and further findings that can be drawn from them (e.g., regarding differential rotation).

\subsection{Spot dominated magnetic activity}

In the past four decades, XX\,Tri essentially became hotter and bluer, resulting in a gradual displacement in the red clump region of the HRD; see Fig.\,\ref{temp_HRD}. This global change is spot-dominated, meaning that huge, dark starspots dominate the variability. When we see more spots, the star is fainter overall, unlike the plage-dominated Sun, whose activity is characterized by the exact opposite: the Sun is brightest at spot maximum. Another characteristic feature of spot dominance is that there is an anti-correlation between the rotational light curve and the chromospheric activity indicators
(e.g. Ca\,{\sc ii}\,H\&K lines or H$\alpha$ emission). We note that this kind of spot dominance is a regular feature of active (sub)giants in long-period RS\,CVn systems; we only mention the most well-known such targets here: II\,Peg and $\lambda$\,And \citep{2008A&A...479..557F,2025A&A...695A..89A}, IM\,Peg \citep{2010PASP..122..670Z}, V711\,Tau \citep{1999ApJS..121..547V}, $\sigma$\,Gem \citep{2001A&A...373..199K}, $\zeta$\,And \citep{2007A&A...463.1071K,2016Natur.533..217R}, etc.

We believe that the significant overall brightness increase of XX\,Tri observed over the past four decades together with the color index changes, which also cause the displacement of the star on the HRD, are due to strong and variable surface magnetic fields present throughout the intensity variations of surface spot activity \citep[for two more examples see][]{2014A&A...572A..94O}. All these changes, with the huge amplitudes in the early period and their decreasing magnitude over time, and the gradual brightening, are difficult to interpret other than that the star's unspotted temperature has also increased over the past four decades. The likely cause of this is the extensive spot activity itself, which blocks the outward energy flux, which eventually, on a longer timescale, breaks through and manifests itself in a global brightening \citep[cf.][see their Discussion section]{2024NatCo..15.9986S}.


\section{Summary and conclusions}\label{sect_sum}

This study is a continuation, the second part of the series of papers that began with the recently published Paper~I, which presented an unprecedented Doppler imaging study of XX\,Tri, visualizing the continuous surface spot evolution for 16 years. In the present `Paper~II' we attempted to perform a comprehensive analysis of the star's four-decade photometric history. 
Our main conclusions drawn from the results are the following:

\begin{enumerate}
      \item The most significant of the longer-term cycles appearing in photometric data series is $\approx$4 years, consistent with the 4.1-year cycle found from independent spectroscopic time series in Paper~I. This mid-term cycle apparently has nothing to do with the star's gradual brightening over several decades. This cycle indicates rather a flip-flop-like behavior. Compared to the time-series Doppler images of Paper~I, it appears that typically during this time the 2-3 active longitudes that are usually present on the stellar surface are rearranged. 
      \item Comparing the seasonal changes in the rotation period with the results of previous Doppler studies suggests that XX\,Tri performs solar-like surface differential rotation, although the surface shear is significantly smaller than that of the Sun. This is further supported by the weak, sun-like differential rotation with a shear coefficient of $\alpha = 0.014$, inferred from a preliminary analysis of the time-series Doppler images from Paper~I. Our result is in good agreement with the empirically predicted value for such giant components in RS\,CVn type close binary systems. 
      \item The average brightness of the star has gradually increased over the four decades of observation, while the spot coverage and the seasonal amplitudes of the brightness variability in general have decreased. However, the observed changes cannot be interpreted solely as changes in the number and size of the spots, we must also assume that the surface temperature, that is the unspotted brightness, has also increased over the decades. This result should prompt users of photometric spot models to reconsider the basic concept of a constant unspotted temperature when interpreting long-term trends in brightness changes of spotted stars.
   \end{enumerate}

\section*{Data availability}

The previously unpublished photometric data of XX\,Tri used in this paper (listed in Sect.\,\ref{sect_obs}) are available in electronic form at the CDS via anonymous ftp to \url{cdsarc.u-strasbg.fr} (\url{130.79.128.5}) or via \url{http://cdsweb.u-strasbg.fr/cgi-bin/qcat?J/A+A/}.

\begin{acknowledgements}
The authors gratefully acknowledge the comments and suggestions of the reviewer, Dr S. Gu, which helped improve the paper. This work was supported by the Hungarian National Research, Development and Innovation Office grant KKP-143986.
GWH acknowledges long-term support from NASA, NSF, and the Tennessee State University.
STELLA was made possible by funding through the State of Brandenburg (MWFK) and the German Federal Ministry of Education and Research (BMBF). The facility is a collaboration of the AIP in Brandenburg with the IAC in Tenerife. 
This work has made use of data from the European Space Agency (ESA) mission {\it Gaia} (\url{https://www.cosmos.esa.int/gaia}), processed by the {\it Gaia} Data Processing and Analysis Consortium (DPAC, \url{https://www.cosmos.esa.int/web/gaia/dpac/consortium}). Funding for the DPAC has been provided by national institutions, in particular the institutions participating in the {\it Gaia} Multilateral Agreement.

\end{acknowledgements}

\bibliography{xxtripaper2}

\begin{appendix}

\section{Lomb-Scargle periodograms}\label{LS_periodograms}

\begin{figure}[h]
    \includegraphics[width=\columnwidth]{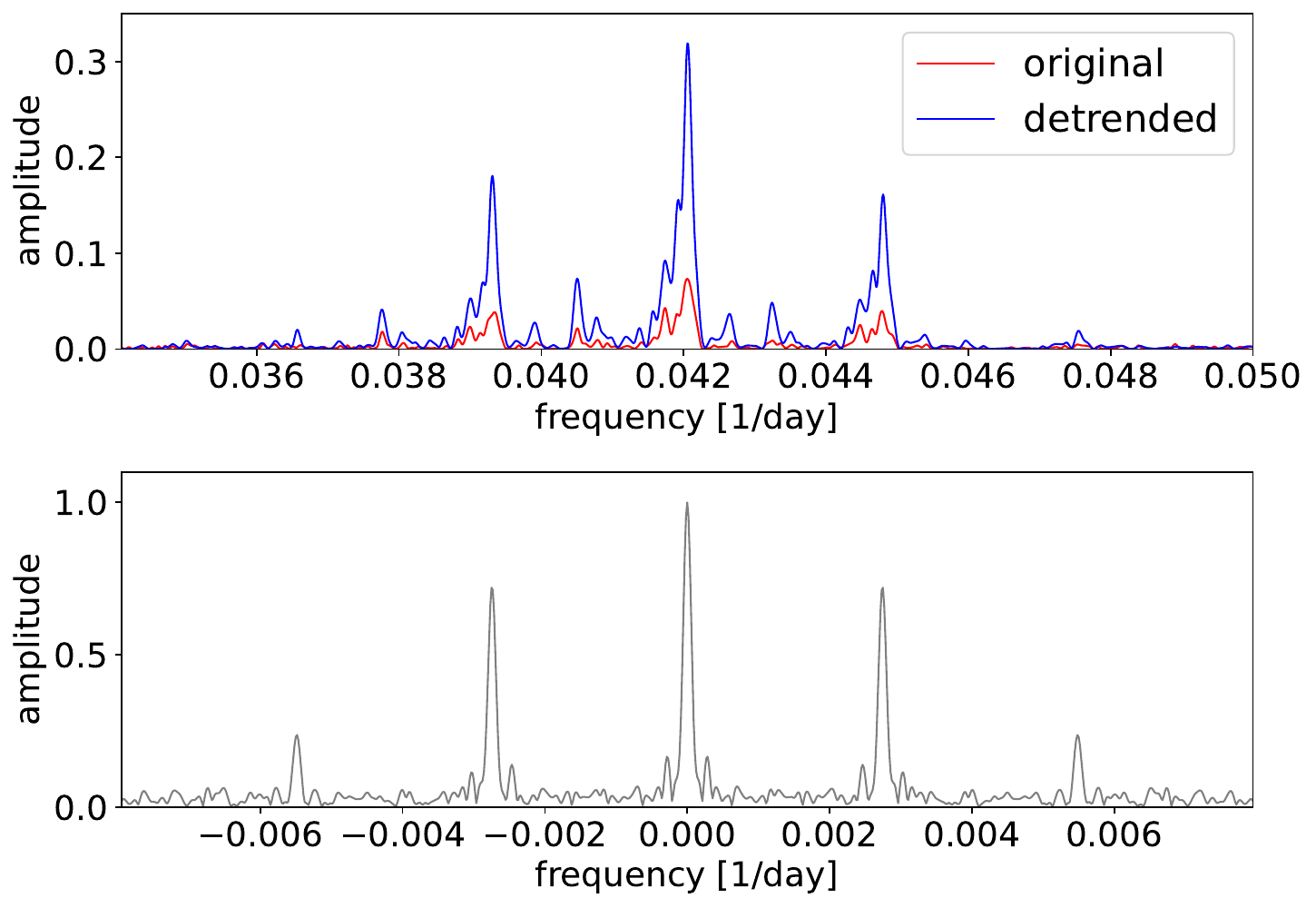}\\
      \caption{Top panel: part of the Lomb-Scargle periodogram for the entire $V$ data showing the rotation-related periods. The highest peak of the main lobe corresponds to the strongest period of $\approx$23.78\,d, the other two large lateral ones (and the other much smaller ones) are side lobes from windowing. Bottom panel: the corresponding window function.}
      \label{LS_full}
\end{figure}

\begin{figure}[h]
    \includegraphics[width=\columnwidth]{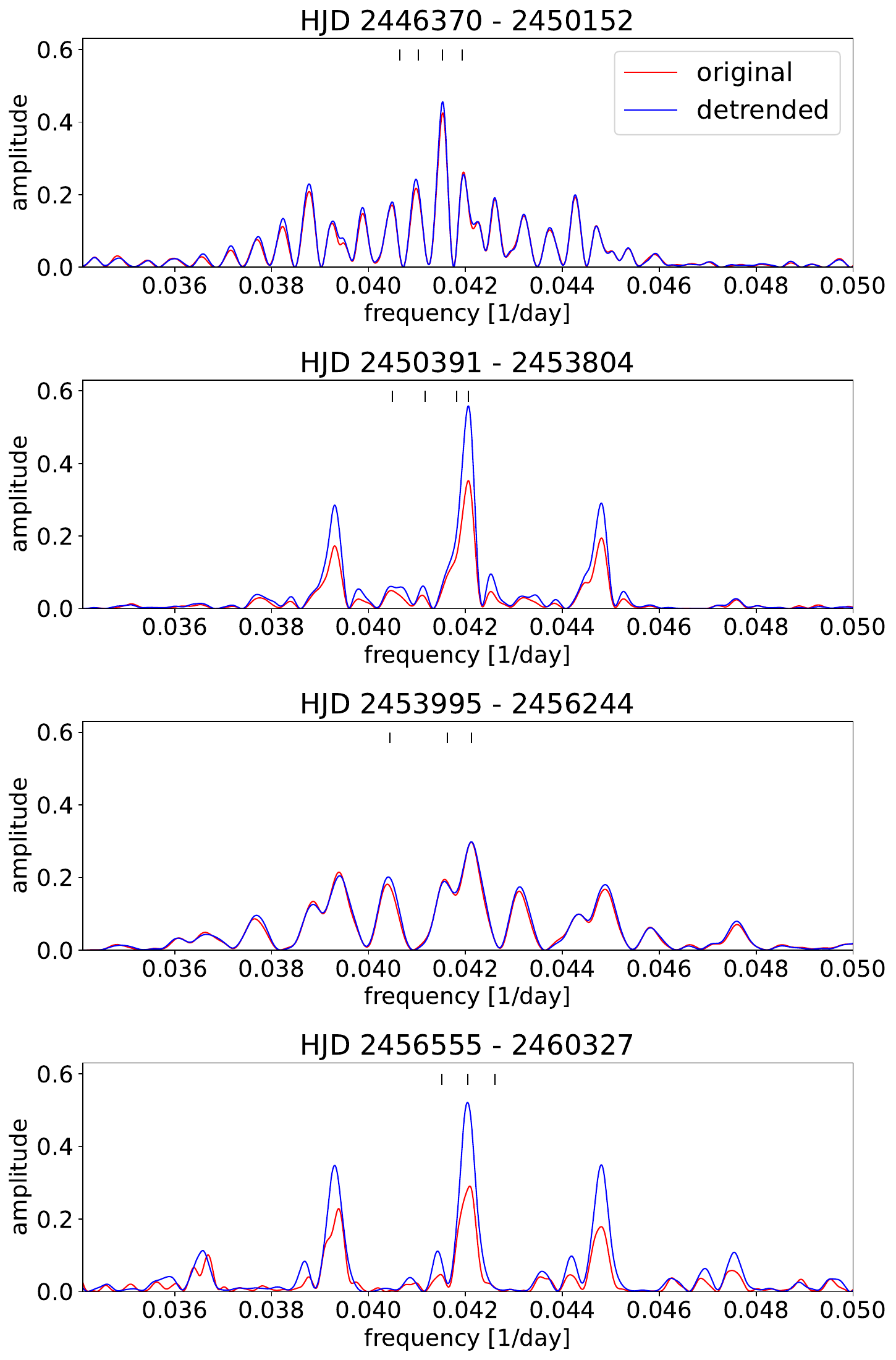}\\
      \caption{The rotation-related parts of the Lomb-scargle periodograms obtained for the four sub-seasons; see also Fig.\,\ref{fig_per-ampl}. The ticks at the top of each panel are the locations of the periods listed in Table\,\ref{tab_periods}.}
      \label{LS_subsets}
\end{figure}

In connection with Fig.\,\ref{fig_per-ampl} and Table\,\ref{tab_periods}, we present here the Lomb-Scargle periodograms obtained for the entire $V$ data set (see Fig.\,\ref{LS_full}) and the four sub-seasons (see Fig.\,\ref{LS_subsets}).
We note that the amplitudes of the periodograms presented here do not necessarily match the final values in Table\,\ref{tab_periods}, as those in the table were derived from the successive uses of the \texttt{MuFrAn} code (see Sect.\,\ref{sect_methods}), in which the next period search was always performed on the spectrum prewhitened by the previously found dominant period. Nevertheless, the locations of the relevant Lomb-Scargle peaks agree well with the corresponding period values in Table\,\ref{tab_periods}.

\end{appendix}

\end{document}